\documentclass[conference,letterpaper]{IEEEtran}

\usepackage{cite}

\usepackage{amsthm}

\newtheorem{theorem}{Theorem}

\usepackage{multirow}
\usepackage{braket}
\usepackage{lineno,xcolor}
\usepackage{amsmath}

\usepackage{amsfonts}%
\usepackage{amssymb}
\usepackage{wasysym}
\usepackage{stmaryrd}
\usepackage{bbm} 
\usepackage{tabularx}
\usepackage{multirow}
\usepackage{mathtools}
\usepackage{enumerate}
\usepackage{subfigure}

\usepackage{pgf}
\usepackage{bm}
\usepackage{multicol}
\usepackage{color}
\usepackage{url}

\usepackage{bibunits}
\usepackage{wrapfig}
\usepackage{sidecap}
\usepackage{soul,xcolor}

\usepackage{eurosym}
\usepackage{mathrsfs}
\usepackage[utf8]{inputenc}

\usepackage{algorithmicx}
\usepackage{algpseudocode}
\usepackage{algorithm}
\usepackage{stmaryrd}
\usepackage{upgreek}
\usepackage[hidelinks]{hyperref}
\usepackage{paralist}
\usepackage{graphicx}        

\usepackage{setspace}
\usepackage{tikz}
\usetikzlibrary{arrows}
\usepackage{cleveref}

\usepackage{subfigure}

\usepackage{etoolbox}
\makeatletter
\patchcmd{\@makecaption}
{\scshape}
{}
{}
{}
\makeatother

\DeclareMathAlphabet{\mathpzc}{OT1}{pzc}{m}{it}

\usepackage{epstopdf}
\usepackage{soul}
\setstcolor{red} 
\setul{0pt}{0.7pt}

\newtheorem{definition}[theorem]{Definition}
\IEEEoverridecommandlockouts

\usepackage{enumitem}
\setlist{nolistsep}
\newcommand*\xor{\oplus}

\begin{document}
\title{Multistage Rewinding Decoder \\ for {QLDPC} Codes}
\author{Milad~Taghipour, Dimitris~Chytas, and Bane~Vasi\'{c},~\IEEEmembership{Fellow,~IEEE}\\
\IEEEauthorblockA{Department of Electrical and Computer Engineering, The University of Arizona, Tucson, AZ.}
Email: \{miladt, dchytas\}@arizona.edu, vasic@ece.arizona.edu
\thanks{The authors acknowledge the support of the National Science Foundation under grants CIF-2420424, CIF-2106189, CCF-2100013. Bane Vasi\'c has disclosed an outside interest in his startup company QEC Labs to The University of Arizona. Conflicts of interest resulting from this interest are being managed by The University of Arizona in accordance
with its policies.}
\thanks{\copyright~2026 IEEE. Personal use of this material is permitted. Permission from IEEE must be obtained for all other uses, in any current or future media, including reprinting/republishing this material for advertising or promotional purposes, creating new collective works, for resale or redistribution to servers or lists, or reuse of any copyrighted component of this work in other works.}
}

\maketitle

\begin{abstract}
In this paper, we propose a multistage decoding framework that leverages internal information produced by an underlying message-passing decoder. The proposed method targets the failure dynamics caused by both classical trapping sets and degenerate errors supported on symmetric stabilizers, which are among the primary limitations of iterative decoding for QLDPC codes. To identify unreliable variable nodes, we introduce a heuristic metric that combines several dynamical features of the decoder, including variable-node log likelihood reliabilities, hard-decision oscillations, the number of adjacent unsatisfied checks, and the soft information contributed by unsatisfied checks. Based on this ranking metric, the decoder performs guided rewinds by selectively forcing the initial log likelihood ratio values of the most suspicious variable nodes and restarting the message-passing decoder under the corresponding forced configuration. To manage the combinatorial growth of candidate configurations, the search is formulated within a beam-search framework with controlled beam width. In addition, we introduce a pruning metric based on the combination of the residual syndrome weight and a posteriori reliability of the decoder output, thereby retaining only the most promising search paths. Logical error rate results demonstrate that the proposed decoder significantly outperforms the normalized min-sum decoder and achieves competitive performance with belief propagation enhanced by order-$10$ ordered statistics decoding.
\end{abstract}

\begin{IEEEkeywords}
QLDPC codes, min-sum decoding, multistage decoding, trapping sets, degeneracy.
\end{IEEEkeywords}
\section{Introduction}
\IEEEPARstart{Q}{uantum} low-density parity check (QLDPC) codes have gained attention due to their high minimum distance and asymptotically good code rate scaling~\cite{panteleev2021quantumLinearMinDLocalTestable, QuantumTannerCodeszemor,vasic2025quantumlowdensityparitycheckcodes}, with extensions to non-binary constructions receiving recent attention~\cite{KentaNonBinary, Spencer2026quditlowdensity, ShantomNonBinary}. Families of non-topological QLDPC codes, such as Bivariate Bicycle (BB) codes, have been shown to outperform topological codes in quantum error correction (QEC)~\cite{bravyi2024high}, albeit at the cost of increased qubit connectivity. 
At the same time, the practical use of QLDPC codes depends critically on the availability of decoders that can reliably handle the failure dynamics that arise in their Tanner graphs. Recent developments suggest that hypergraph-product (HP) and lifted-product (LP) codes are promising for neutral-atom architectures~\cite{xu2024constant}, while BB codes are natural candidates for superconducting platforms~\cite{bravyi2024high}. These developments further motivate the design of decoders that can overcome the trapping sets that limit the performance of iterative decoding of QLDPC codes. 
Iterative message-passing decoders, such as belief propagation (BP) and min-sum (MS), are attractive because of their strong performance for classical LDPC codes. However, their performance on QLDPC codes is often limited by harmful subgraph configurations.

The first class consists of classical trapping sets. These are the same type of harmful configurations that appear in classical LDPC decoding, where unreliable messages reinforce one another and prevent convergence~\cite{ontology,03Richardson}. In classical LDPC decoding, message-passing performance is strongly tied to the locally tree like structure of the Tanner graph. By contrast, many QLDPC constructions contain multiple short cycles and other small subgraph structures induced by their low weight stabilizers and quantum commutativity constraints. As a result, message-passing decoders can struggle to produce reliable local inference. This issue is especially relevant for product code constructions such as HP and LP codes, whose Tanner graphs are built from the Tanner graphs of constituent classical LDPC codes~\cite{pradhan2025lineartimeiterativedecoders}. Consequently, trapping sets present in the constituent codes can reappear as multiple isomorphic copies in the Tanner graph of the resulting QLDPC code. We refer to these inherited configurations as classical trapping sets.

The second class consists of configurations intrinsic to the quantum setting. Unlike classical error correction, where the decoder aims to recover the exact error pattern, QEC only requires recovery up to a stabilizer. Thus, multiple distinct error patterns can produce the same syndrome and correspond to the same logical action on the encoded state. This phenomenon, known as \textit{degeneracy}, creates ambiguities that standard message-passing decoders are not designed to resolve. In particular, degeneracy becomes increasingly relevant when the minimum distance grows relative to the stabilizer weight~\cite{roffe_decoding_2020}. Failures caused by such stabilizer induced ambiguities can often be traced to subgraphs supported on stabilizers or combinations of stabilizers; these are commonly referred to as quantum trapping sets, or stabilizer induced trapping sets~\cite{quantumTS}. Therefore, decoding QLDPC codes requires addressing both classical trapping sets inherited from LDPC structure and quantum trapping sets induced by degeneracy.

Several approaches have been proposed to improve iterative decoding performance on QLDPC codes. Probabilistic and deterministic modifications of BP and two bit flipping~\cite{Poulin_2008, DimitrisEnhanced, BPOTS, ChytasCollective}, as well as recent variants such as Relay-BP~\cite{muller2025improvedbeliefpropagationsufficient}, have shown improved performance by modifying the message-passing dynamics, exploiting degeneracy, or using disordered memory strengths to damp oscillations and break symmetries. However, these methods are primarily designed to improve the BP dynamics itself and do not explicitly use a trapping set oriented reliability metric to target both classical trapping sets inherited from the Tanner graph structure and quantum trapping sets induced by degeneracy. Scheduling strategies, such as layered BP and MS algorithms, can further enhance decoding by leveraging degeneracy, albeit with added latency~\cite{layered}. 

Post processing methods, including ordered statistics decoding (OSD)~\cite{osd}, stabilizer inactivation (SI)~\cite{StabInactivation_Julien_2022}, and BP with guided decimation (BPGD)~\cite{decim, 11154483, window}, further improve performance at the cost of increased complexity, e.g., $O(n^3)$, $O(n^2\log n)$, and $O(n^2)$ for block length $n$, respectively. Recent techniques, such as~\cite{amb,local}, improve the efficiency of BP-OSD's matrix inversion. Other post processing methods tailored to accurate noise models, such as BP plus ordered Tanner forest (BP+OTF)~\cite{otf} and the SymBreak decoder~\cite{symbreak}, have also been introduced. These methods typically rely on reliability information obtained from the iterative decoder, such as a posteriori log likelihood ratios or related soft output quantities, to guide post processing decisions. While effective, such reliability measures may not fully capture the trapping sets responsible for decoder failures. In a related direction, beam-search and related search based decoding methods have also been explored for quantum LDPC decoding, including BP-guided beam search decoders and most likely error search decoders~\cite{ye2025beamsearchdecoderquantum, beni2025tesseractsearchbaseddecoderquantum}.

In the classical decoding literature, several related strategies have been explored. These include multiple decoding attempts with random initializations~\cite{MUDRI_DECODER_ICC_2015}, the use of guided information from suspicious variables to enable rewinding in the GDBF and BP decoders~\cite{suspicion_propagation, AugmentedBP}, the incorporation of randomness into variable node and check node update rules to improve decoding performance and escape from trapping sets~\cite{NGBFSundar, one_error_drives_another_tcomm_2015}, as well as beam search optimization techniques for reinforcement learning aided decoding~\cite{ActionListRL, taghipour2025actionlistreinforcementlearningsyndrome}. 

In this work, we employ message-passing decoders iteratively within a multistage framework to mitigate the effect of trapping sets present in the Tanner graph of the code. To this end, we define an unreliability metric based on internal decoder information, including hard-decision oscillations, the connectivity of unsatisfied checks to variable nodes, the opposing soft check-to-variable messages associated with unsatisfied checks, and variable node log likelihood reliabilities. This heuristic metric combines these sources of information into a single criterion for identifying unreliable variable nodes. By ranking the variable nodes according to the proposed metric, we identify those that are most likely to be erroneous and responsible for preventing the decoder from converging. Based on this ranking, we then apply a multistage decoding procedure in which the values of the identified variable nodes are selectively forced. This process is carried out in a beam search framework~\cite{zhou2005beam}, starting from the root node corresponding to the measured syndrome. At each stage, unreliable variable nodes are identified using the proposed metric, their initial log likelihood ratios (LLRs) values are forced, and the message-passing decoder is rewind under the corresponding forced variable configuration. Since the number of candidate paths in the search tree can grow rapidly with the search depth, we introduce a beam width parameter to control the search complexity and ensure practical feasibility. In addition, we define a pruning metric based on a combination of the a posteriori log likelihood ratios and the corresponding residual syndrome of the decoder, which is used to retain the most promising search paths at each stage.

We focus on two-block Calderbank-Shor-Steane (CSS) codes and lifted-product codes. In particular, we consider four BB codes with parameters $[[72, 12, 6]]$, $[[108, 8, 10]]$, $[[144, 12, 12 ]]$, and $[[288, 12, 18]]$, as well as the lifted-product Tanner code $[[1054, 124, 20]]$. Across these code families, the proposed method yields consistent performance improvements. In particular, for the BB codes, the proposed decoder is able to match and in some cases surpass the performance of BP-OSD. For example, at crossover probability $\alpha = 0.03$, the proposed multistage decoder reduces the logical error rate by two orders of magnitude for the $[[288,12,18]]$ code. Similarly, at crossover probability $\alpha = 0.04$, it achieves two orders of magnitude improvement for the lifted product Tanner code $[[1054,124,20]]$ code. More detailed discussion of the simulation results is provided in Section~\ref{sec:performance}.

The rest of the paper is organized as follows. In Section~\ref{sec:pre}, we introduce the preliminaries of QEC and provide an overview of syndrome MS decoding. In Section~\ref{sec:TS}, we discuss the trapping sets and identification of unreliable variable nodes. In Section~\ref{sec:MultiStage}, we present the proposed multistage decoding algorithm. Finally, in Section~\ref{sec:performance}, we provide simulation results for BB codes and the LP Tanner code.

\section{Preliminaries}
\label{sec:pre}
\subsection{Stabilizer Formalism}
Let \(\mathbb{F}_2^n\) denotes the $n$-dimensional vector space over \(\mathbb{F}_2\). The \textit{Hamming weight} of an element in \(\mathbb{F}_2^n\) is defined as the number of its non-zero entries. An \([n, k, d]\) linear code \(C \subset \mathbb{F}_2^n\) is a linear subspace of \(\mathbb{F}_2^n\) spanned by \(k\) independent vectors, with a minimum Hamming weight of at least \(d\). A code \(C\) can also be represented by a parity check matrix \(H\), where \(C = \ker H\). For simplicity, we consider regular parity check matrices, with column-weight \(d_v\) and row-weight \(d_c\).

Let \((\mathbb{C}^2)^{\otimes n}\) denotes the \(2^n\)-dimensional Hilbert space of $n$ qubits, and \(P_n\) the \(n\)-qubit Pauli group. A \textit{stabilizer} group is an Abelian subgroup \(S \subset P_n\) not containing $-I$. An \(\llbracket n,k,d \rrbracket\) stabilizer code is a \(2^k\)-dimensional subspace \(\mathcal{C} \subset (\mathbb{C}^2)^{\otimes n}\) such that 
\[
s_i\ket{\Psi} = \ket{\Psi}, \quad \forall\ s_i \in S, \ \ket{\Psi} \in \mathcal{C}.
\]
A \(\llbracket n, k_X - k_Z^{\perp}, d \rrbracket\) CSS code \(\mathcal{C}\) is a type of stabilizer code constructed using two classical codes \(C_X = \ker H_X\) and \(C_Z = \ker H_Z\), satisfying \(C_Z^{\perp} \subset C_X\) and \(C_X^{\perp} \subset C_Z\)~\cite{calderbank1996quantum_exists}. The minimum distance \(d = \min\{d_X, d_Z\}\) is determined by the Hamming weights of codewords in \(C_X \setminus C_Z^{\perp}\) and \(C_Z \setminus C_X^{\perp}\). For additional details, refer to~\cite{Gottesman97}.

CSS codes allow binary decoding, i.e., $X$ errors are decoded using $H_Z$, and $Z$ errors are decoded using $H_X$. Thus, two independent binary symmetric channels (BSCs) can be considered~\cite{mackay_quantum}. Let us denote as $\mathbf{e}=[\mathbf{e}_X, \mathbf{e}_Z]$ the binary representation
of a Pauli error acting on the $n$ qubits. The corresponding input syndromes are obtained as $\mathbf{s}_Z= \mathbf{e}_X  H_Z^T$
and $\mathbf{s}_X= \mathbf{e}_Z  H_X^T$, or in a compact form, $\mathbf{s}=[\mathbf{s}_X, \mathbf{s}_Z]$. 

A zero syndrome implies no detectable errors, while a non-zero syndrome detects errors. Syndrome-based decoders estimate \(\hat{\mathbf{e}}\) such that \(\mathbf{\hat{s}} = \mathbf{s}\). Decoding succeeds if \(\mathbf{e} \oplus \hat{\mathbf{e}}\) belongs to the rowspace of \(H\). Failure occurs if the syndrome doesn't match or a logical error results, where \(\mathbf{e} \oplus \hat{\mathbf{e}}\) commutes with all stabilizers but it is not in the rowspace of $H$.

The $H_X$ and $H_Z$ stabilizer matrices can each be represented as a \textit{Tanner graph},  which is a bipartite graph with two sets of nodes:  $n$ variable (qubit) nodes denoted by $j \in \{1, 2, ...,n\}$ and  $m$ check nodes denoted by $i \in \{1, 2, ...,m\}$. If the $(i,j)$-th entry of the binary parity check matrix is not zero, then there is an edge connecting the variable node $j$ and check node $i$ (edges connect neighboring nodes).
We denote by $\mathcal{M}(j)$ the indices of the neighboring check nodes of the variable node $j$, and by $\mathcal{N}(i)$ the indices of the neighboring variable nodes of the check node $i$.
For the rest of this paper, only one type of error ($X$ error) and its decoding will be considered; without loss of generality, the notation $H$ will refer to $H_Z$, $\mathbf{e}$ will refer to $\mathbf{e}_X$, and $\mathbf{s}$ will refer to $\mathbf{s}_Z$. Next we will describe the two families of CSS codes we consider in this paper.

\subsection{Generalized and Bivariate Bicycle Codes}
In this work, we consider two-block CSS codes, a practically important family of QLDPC codes that includes Bicycle codes~\cite{mackay_quantum}, generalized Bicycle (GB) codes, and Bivariate Bicycle (BB) codes~\cite{bravyi2024high,PhysRevA.109.022407}. Compared with more general constructions such as lifted-product codes, BB/GB codes retain a particularly simple two-block structure, which makes them attractive for finite-length designs and hardware-oriented implementations. In particular, BB codes have received significant attention because their Tanner graphs admit a highly structured geometric layout, enabling embeddings in two interconnected planar layers and making them promising candidates for superconducting-qubit architectures~\cite{bravyi2024high}.

To describe this family, let \(A\) and \(B\) be two commuting binary \(n\times n\) matrices, i.e., \(AB=BA\). The CSS parity check matrices are defined as
\begin{equation}
H_X=[A,B], \qquad H_Z=[B^T,A^T].
\label{eq:pcm}
\end{equation}
The commutativity condition guarantees
\[
H_XH_Z^T = AB+BA = 0,
\]
so the CSS orthogonality constraint is satisfied. The original Bicycle codes of~\cite{mackay_quantum} arise as a special case, typically with circulant blocks. BB codes form a more structured subclass in which \(A\) and \(B\) are expressed as sums of monomials in two commuting shift operators, giving rise to quasi-cyclic codes with strong algebraic structure and favorable implementation properties. 

\subsection{Lifted-Product Codes}
Compared with BB codes, lifted-product (LP) codes provide a more general product-based construction of QLDPC codes and have been shown to achieve dimension scaling on the order of \(\log n\) and distance scaling on the order of \(n/\log n\), improving on the minimum-distance scaling of hypergraph-product codes~\cite{panteleev2020quantumLinearMinD}. While BB codes arise from a two-block CSS structure with commuting constituent matrices, LP codes are obtained by combining two sparse classical codes through a product construction, which provides greater design flexibility while still yielding sparse, highly structured QLDPC codes. In the quasi-cyclic setting, the construction is described through base matrices in exponent form, whose entries specify powers of an \(L\times L\) circulant permutation matrix, with \(-\infty\) denoting the all-zero block. Replacing each exponent by the corresponding circulant block produces the lifted binary parity check matrices.

More specifically, given two quasi-cyclic base matrices \(B_1\in\mathbb{Z}_L^{m_1\times n_1}\) and \(B_2\in\mathbb{Z}_L^{m_2\times n_2}\), the LP construction defines
\[
B_X=[\,B_1\otimes I_{n_2}\;\; I_{m_1}\otimes B_2^\star\,], \qquad
B_Z=[\,I_{n_1}\otimes B_2\;\; B_1^\star\otimes I_{m_2}\,],
\]
where \(\otimes\) denotes the Kronecker product and \(B^\star\) is the conjugate transpose of the exponent matrix. After lifting, one obtains binary parity check matrices \(H_X=\mathrm{L}(B_X)\) and \(H_Z=\mathrm{L}(B_Z)\), which satisfy \(H_XH_Z^T=0\) and therefore define a CSS code. In this work, we will focus on the $[[1054,124,20]]$ LP code formed by the classical Tanner code of length $155$~\cite{raveendran2025minimumdistancesfinitelengthlifted}, which we will refer to as LP Tanner code.

\subsection{Syndrome based MS decoding}
\label{sec:MS}
MS decoding is a low-complexity variant of BP decoding. Normalized min-sum (nMS) provides a good approximation of BP decoding if the normalization parameter is carefully selected~\cite{05CDEFH}. Syndrome-based MS has been applied for QEC, but good results are only obtained when post processing techniques like OSD and SI are applied, or when serial/adaptive scheduling is used~\cite{layered,memory}. 
We will now briefly describe the syndrome-based nMS decoder. 

A check-to-variable message sent from check node $i$ to variable node $j$ at the $\ell$-th iteration is denoted by $\mu^{(\ell)}_{i,j}$, whereas a variable-to-check message sent from variable node $j$ to check node $i$ at the $\ell$-th iteration is denoted by $\nu^{(\ell)}_{j,i}$. Assuming a noise-free syndrome measurement and a binary symmetric channel inducing only $X$ errors with probability $\alpha$, the measured syndrome value is $s_i \in \{0,1\}$, $i \in \{1, 2, ..., m\}$ and the a priori log likelihood ratio value for every variable node is given by $\lambda_j=\lambda=\log({\frac{1-\alpha}{\alpha}})$. For the sake of brevity, we denote the natural numbers from  $1$ to $n$ by $[n]$.
    
The variable node update rule is computed as: 
\begin{equation}
    \nu^{(\ell)}_{j,i} = \lambda_j + \sum\limits_{i' \in \mathcal{M}(j) \backslash \{ i \}}\mu^{(\ell)}_{i',j},
    \label{eq:vnu}
\end{equation}
where $\nu^{(0)}_{j,i}=\lambda_j$, and the check node update rule is computed as: 
\begin{equation}
   \mu^{(\ell)}_{i,j} = (1-2s_i)\prod\limits_{j' \in \mathcal{N}(i) \backslash \{j\}} \mathrm{sign}(\nu^{(\ell-1)}_{j',i})\min\limits_{j' \in \mathcal{N}(i) \backslash \{j\}}|\nu^{(\ell-1)}_{j',i}|,
    \label{eq:cnu}
\end{equation}
where $\mathrm{sign}$ is the sign function. 

In this work, we use the normalized version of MS algorithm. Hence, check-to-variable messages are multiplied by a scalar $\beta$.
We define the a posteriori probability (APP) as
\begin{equation}
    \zeta_j^{(\ell)} = \lambda_j + \sum\limits_{i \in \mathcal{M}(j) } \mu^{(\ell)}_{i,j}.
\end{equation}
Finally, the error estimate $\hat{\mathbf{e}}^{(\ell)}$ at the $\ell$-th iteration is given by the decision update function:
\begin{equation}
\hat{e}_j^{(\ell)}=
\begin{cases}
    0, & \zeta_j^{(\ell)} > 0,\\
    1, & \zeta_j^{(\ell)} \le 0.
\end{cases}
\label{eq:dec}
\end{equation}
Let $\mathbf{s}^{\mathrm{tar}} $ denote the measured syndrome, and let $\mathbf{s}^{\mathrm{dec},\ {(\ell)}} $ denote the decoder syndrome at iteration $\ell$ induced by the estimated error pattern $\hat{\mathbf{e}}^{(\ell)}$, i.e.,
\begin{equation}
    \mathbf{s}^{\mathrm{dec}, \ {(\ell)}} = \hat{\mathbf{e}}^{(\ell)} H^T.
\end{equation}
We define the residual syndrome by
\begin{equation}
    \mathbf{s}^{\mathrm{res}} = \mathbf{s}^{\mathrm{dec}, \ (\ell)} \xor \mathbf{s}^{\mathrm{tar}}.
\end{equation}
The decoding procedure is continued until the maximum number of iterations $L$ has been reached or until the residual syndrome at the $\ell$-th iteration equals the all-zero vector.

\section{Trapping Sets and Subgraphs}
\label{sec:TS}
Due to the presence of cycles in the Tanner graph, the belief propagation algorithm is generally suboptimal. In this section, we examine the uncorrectable configurations known as trapping sets~\cite{ontology} and then introduce a metric for estimating the locations of unreliable variable nodes. After predefined number of iterations for the syndrome based iterative decoding, the
decoding is declared unsuccessful for a given input syndrome if the decoder is not able to
identify an error pattern whose syndrome matches to the input syndrome. Equivalently, a decoding failure occurs whenever the residual syndrome is not the all-zero vector, i.e., $\mathbf{s}^{\mathrm{res}} \neq \mathbf{0}$. To characterize the problematic configuration for the iterative decoding we introduce the following notation. A check node $i$ is said to be eventually satisfied if there exists a positive integer $\ell^{\prime}$ such that $s^{\mathrm{res}}_i = 0, \ \forall \ell \geq \ell^{\prime} $. A variable node $j$ is said to be eventually converged if there exists a positive integer $\ell^{\prime}$ such that $\hat{e}_j^{(\ell+1)} = \hat{e}_j^{(\ell)}, \ \forall \ell \geq \ell^{\prime}$. Using these notions, a quantum trapping set is defined as follows.
\begin{definition}[Trapping Set~\cite{quantumTS}]
    A trapping set $\mathcal{TS}$ for a syndrome based iterative decoder is a non-empty set of variable nodes in a Tanner graph that either fail to eventually converge or are connected to check nodes that fail to become eventually satisfied. If the induced subgraph contains $a$ variable nodes and $b$ odd-degree check nodes, then $\mathcal{TS}$ is labeled as an $(a, b)$ trapping set.
\end{definition}

In general, trapping sets are detrimental configurations that can cause iterative decoding to fail. In the following, we introduce a metric to identify variable nodes that are likely to belong to such configurations. 
To rank variable nodes according to their likelihood of belonging to the residual error pattern, we define a candidate-selection metric $M_j$ for each variable node $j$ as
\begin{equation}
    M_j = \frac{N_j}{D_j + \epsilon},
    \label{eq:metric}
\end{equation}
Where $\epsilon > 0$ is a small constant introduced for numerical stability. The numerator $N_i$ is a weighted suspiciousness score given by
\begin{equation}
N_j =
    c_U \frac{U_j}{\max_k U_k + \epsilon} +
    c_E \frac{E_j}{\max_k E_k + \epsilon} +
    c_O \frac{O_j}{\max_k O_k + \epsilon},
    \label{eq:num}
\end{equation}
and the denominator $D_j$ is a normalized reliability term defined as
\begin{equation}
    D_j = \frac{|\zeta_j|}{\max_k |\zeta_k| + \epsilon}.
\end{equation}
Here, $c_U$, $c_E$, and $c_O$ are nonnegative weighting coefficients that control the relative contribution of the residual unsatisfied check participation, the opposing check-to-variable message relative to the final APP sign, and the temporal oscillation, respectively.
The first component in Equation \ref{eq:num},
\begin{equation}
    U_j = \sum_{i \in \mathcal{N}(j)} s^{\textrm{res}}_i,
\end{equation}
counts the number of residual unsatisfied checks incident to variable node $j$. Thus, $U_j$ provides a suspicious metric: a variable node connected to many unsatisfied parity checks is more likely to be involved in the local configuration responsible for decoding failure.
The second component in Equation \ref{eq:num},
\begin{equation}
    E_j = \sum_{i \in \mathcal{N}(j)} s^{\textrm{res}}_i \max\bigl(0, -\operatorname{sign}(\zeta_j)\mu_{i,j}\bigr),
\end{equation}
quantifies the degree to which incoming check-to-variable messages from residual unsatisfied checks oppose the final a posteriori probability (APP) decision. More precisely, $E_j$ accumulates only the magnitudes of those incoming messages whose signs contradict the sign of the final APP value $\zeta_j$. Consequently, this term captures not only whether a variable node participates in unsatisfied checks, but also whether those checks actively provide evidence against the decoder’s final decision. It therefore serves as a more refined indicator than mere check participation.
The third component in Equation \ref{eq:num}, denoted by $O_j$, measures temporal instability during decoding. Specifically, 
$O_j$ is defined as the number of times the hard decision associated with variable node $j$ changes over the iterative decoding process. A large value of $O_j$ indicates that the decoder has difficulty stabilizing the estimate of that variable node, which is often symptomatic of local trapping behavior or weak convergence. Hence, $O_j$ provides dynamical information that is not reflected solely by the final decoder state. The role of the denominator $D_j$ is to discount variable nodes whose final APP magnitudes remain large, and are therefore relatively reliable, even if they are structurally involved in unsatisfied checks. Accordingly, the metric $M_j$ becomes large when variable node $j$ simultaneously satisfies four properties: it is connected to many residual unsatisfied checks, it receives strong contradictory check node evidence, it exhibits oscillatory behavior during decoding, and its final APP magnitude is relatively small. Thus, $M_j$ is designed to assign high value to variable nodes that are both strongly implicated in the decoding failure and weakly supported by the final decoder state. The candidate variables are then selected by sorting $\{ M_j \}_{j=1}^{n}$ in descending order. The resulting ordered index set provides a prioritized list of variable nodes for subsequent forced assignments in the staged beam-search procedure. In this manner, the proposed metric yields a single scalar ranking criterion that integrates contribution of unsatisfied checks, soft opposing check-to-variable message to the final APP, and dynamical indicators of decoding failure. In the next section, we describe the proposed multistage decoding procedure.

\section{Multistage Decoding}
\label{sec:MultiStage}
In this section, we investigate a multistage decoding framework that iteratively exploits the information generated by belief propagation (BP) and integrates it with a guided restart mechanism for BP. We define a search node $v \in \mathcal{T}$, where $\mathcal{T}$ denotes the search tree. At stage $t$, each node $v$ is associated with a forced set consisting of $t$ forced variable nodes together with their assigned values:
\begin{equation}
    \mathcal{F} (v) = \{(j_1, a_1), (j_2, a_2), \ldots, (j_t, a_t) \},
    \ \ a_{\ell} \in \{+A, -A\},
\end{equation}
where $A$ denotes the magnitude of the forced value.
At node $v$, we run the forced normalized min-sum decoder, denoted by
\begin{equation}
    \mathcal{D}_{\mathrm{out}} (v) = \operatorname{FNMS} \bigl( H, s^{\mathrm{tar}}, \mathcal{F}(v) \bigr)
\end{equation}
The decoder output is given by
\begin{equation}
    \mathcal{D}_{\mathrm{out}}(v) = \bigl( \hat{e}(v), s^{\mathrm{dec}}(v), \zeta(v), C(v), o(v) \bigr),
\end{equation}
where $\hat{e} (v)$ is the estimated error, $s^{\mathrm{dec}}(v)$ is the decoded syndrome, $\zeta (v)$ denotes the vector of APP values, $C(v)$ represents the final check-to-variable message at search node $v$, and $o(v)$ denotes the oscillation information associated with node $v$. Accordingly, each search node is represented as
\begin{equation}
    v = \bigl( \mathcal{F}(v), \mathcal{D}(v), \mathcal{P}(v), \operatorname{depth} (v) \bigr).
\end{equation}
where $\mathcal{P}(v)$ denotes the pruning score assigned to node $v$, and $\operatorname{depth} (v)$ denotes its depth in the search tree.

Next, We define the top-$K$ candidate variables at node $v$ as
\begin{equation}
    \Pi_{K} (v) = \{ j_1^{\star} (v), \ldots, j_{K}^{\star} (v) \},
    \label{eq:top-K}
\end{equation}
obtained by sorting the unreliability metric $M_j(v)$, as discussed in Eq.~\ref{eq:metric}, over all currently unforced variables. The set of children of node $v$ is then defined as
\begin{equation}
\begin{aligned}
    \operatorname{ch} (v) = \{ v_{j, a} | \mathcal{F} (v_{j,a}) &= \mathcal{F}(v) \cup \{ (j, a)\ \}, \ j \in \Pi_K(v), \\ & a \in \{ +A, -A  \}  \}.
\end{aligned}
\end{equation}
Thus, each candidate variable generates two child nodes, and therefore $|\operatorname{ch}(v)| \leq 2K$. At stage $t$, the active beam is defined as
\begin{equation}
    \mathcal{B}^{(t)} = \{ v_1^{(t)}, \ldots, v_{w_t}^{(t)} \},
\end{equation}
where $w_t \leq W$, and $W$ denotes the beam width of the search.
The pooled set of all children generated from the current beam is
\begin{equation}
    \operatorname{pc}^{(t+1)} = \bigcup_{v \in \mathcal{B}^{(t)}} \operatorname{ch} (v).
\end{equation}
This pooled set is used to select the best nodes according to the pruning criterion. Specifically, the pruning step is defined as
\begin{equation}
    \mathcal{B}^{(t+1)} = \operatorname{Top}_{W}\bigl(\operatorname{pc}^{(t+1)}, \mathcal{P} \bigr),
\end{equation}
where $\operatorname{Top}_W \bigl(\cdot , \mathcal{P} \bigr)$ returns at most $W$ nodes from $\operatorname{pc}^{(t+1)}$ with the highest pruning scores according to the metric $\mathcal{P}$.

To evaluate the quality of a search node, we define the residual syndrome vector at node $v$ by
\begin{equation}
    \mathbf{s}^{\mathrm{res}} (v) = \mathbf{s}^{\mathrm{dec}} (v) \xor \mathbf{s}^{\mathrm{tar}} (v),
\end{equation}
and its Hamming weight by
\begin{equation}
    \textrm{w}_s (v) = ||  \mathbf{s}^{\textrm{res}} (v)  ||_{1}.
\end{equation}
The residual syndrome hamming weight measures the extent to which the decoder output at node $v$ fails to match the target syndrome. Therefore, a smaller value of $\operatorname{wt}_s(v)$ corresponds to a more promising search node. To quantify overall reliability of the decoder output, we define the average APP strength term as 
\begin{equation}
    \xi (v) = \dfrac{1}{n} \sum_{j=1}^n |\zeta_j (v)| 
\end{equation} 
where $\zeta_j(v)$ denotes the APP value associated with variable node $j$ at search node $v$. This quantity reflects the average confidence of the decoder at node $v$ across all variable nodes. Based on these quantities, we define a pruning score of node $v$ as
\begin{equation}
    \mathcal{P}(v) = -\lambda_s \ \textrm{wt}_s (v) + \lambda_{\xi} \xi (v).
\end{equation}
where $\lambda_s$ and $\lambda_{\xi}$ are nonnegative hyperparameters. Hence, nodes with smaller residual syndrome weights and larger average APP magnitudes receive higher pruning scores and are therefore more likely to be retained in the beam. At each stage of the beam search, the pooled candidate set is pruned by retaining only the top $w$ nodes with the highest pruning scores.

A node $v$ is flagged successful if and only if its residual syndrome is zero, i.e., $\mathbf{s}^{\textrm{res}} (v) = \mathbf{0}$. Thus, the set of successful nodes at stage $t$ is defined as
\begin{equation}
    \mathcal{G}^{(t)} = \{ v \in \operatorname{pc}^{(t)} | \mathbf{s}^{\textrm{res}} (v) = \mathbf{0} \}.
\end{equation}
The search terminates as soon as the first stage $t$ is reached for which $\mathcal{G}^{(t)} \neq \emptyset$. In other words, once at least one successful decoding is identified at the current stage, the algorithm does not proceed to deeper stages. When multiple successful nodes are present in $\mathcal{G}^{(t)}$, the final output is selected according to the minimum Hamming weight of the estimated error pattern. Specifically, the selected node is given by
\begin{equation}
    v^{\star} = \arg \min_{v \in \mathcal{G}^{(t)}} w_H (\hat{\mathbf{e}} (v)).
\end{equation}
Equivalently, among all successful nodes in the terminating procedure, the algorithm chooses the one whose estimated error pattern has the smallest Hamming weight. If no successful node is found up to the prescribed maximum stage $T_{\textrm{max}}$, the algorithm terminates without success. The overall procedure is summarized in Algorithm ~\ref{alg:BeamSearch_algo}. In the following section, we present the performance improvements achieved by the proposed decoder and the associated metrics.

\begin{algorithm}
    \caption{Multistage Forced Normalized Syndrome Min-Sum Decoding}
    \label{alg:BeamSearch_algo}
    \textbf{Input:} Parity check matrix $H$, target syndrome $\mathbf{s}^{\mathrm{tar}}$, channel crossover probability $\alpha$,
    maximum decoder iterations $I_{\max}$, normalization factor $\beta$, maximum search stage $T_{\max}$, beam width $W$, top-$K$ candidate count $K$, forced value $|A|$. \\
    \textbf{Output:} estimated error $\hat{\mathbf{e}}^\star$, residual syndrome $\mathbf{s}^{\mathrm{res}} (v^\star)$.
    \begin{algorithmic}[1]
        \State $L \gets \log\!\bigl(\frac{1-\alpha}{\alpha}\bigr)$
        \State Construct root node $\nu_{\emptyset}^{(0)}$ with $\mathcal{F}(\nu_{\emptyset}^{(0)})=\emptyset$

        \State Run
        \[
        \mathcal{D}\!\left(\nu_{\emptyset}^{(0)}\right)
        =
        \operatorname{FNMS}\!\left(H,\mathbf{s}^{\mathrm{tar}},\mathcal{F}\!\left(\nu_{\emptyset}^{(0)}\right)\right)
        \]
        \If{$\mathbf{s}^{\mathrm{res}}(v) = \mathbf{0}$}
            \State \Return $\hat{\mathbf{e}} (v)$
        \EndIf

        \State $\mathcal{B}^{(0)} \gets \{v_{\emptyset}^{(0)}\}$

        \For{$t=0,1,\dots,T_{\max}-1$}
            \State $\operatorname{ch}^{(t+1)} \gets \emptyset$ \Comment{pooled children at stage $t+1$}
        
            \ForAll{$\nu \in \mathcal{B}^{(t)}$}
                \State Compute the top-$K$ candidate variables $\Pi_K(v)$ as described in Equation \ref{eq:top-K}.
                \ForAll{$i \in \Pi_K(v)$}
                    \ForAll{$a \in \{+A,-A\}$}
                        \State Create child node $\nu'$ with
                        \[
                        \mathcal{F}(\nu') \gets \mathcal{F}(\nu)\cup\{(i,a)\}
                        \]
                        \State Run
                        \[
                        \mathcal{D}(\nu')
                        =
                        \operatorname{FNMS}\!\left(H,\mathbf{s}^{\mathrm{tar}},\mathcal{F}(v')\right)
                        \]
                        \State Compute residual syndrome $\mathbf{s}^{\textrm{res}} (v{'})$.
                        \State Compute pruning score in Equation $\mathcal{P} (v{'})$.
                        \State $\operatorname{pc}^{(t+1)} \gets \operatorname{pc}^{(t+1)} \cup \{v' \}.$
                    \EndFor
                \EndFor
            \EndFor
        
            \State $\mathcal{G}^{(t+1)} \gets \{v \in \operatorname{pc}^{(t+1)} | \mathbf{s}^{\textrm{res}} (v) =\mathbf{0} \}$
        
            \If{$\mathcal{G}^{(t+1)} \neq \emptyset$}
                \State Select
                \[
                \nu^\star
                \gets
                \arg\min_{\nu \in \mathcal{S}^{(t+1)}}
                w_H\!\bigl(\hat{\mathbf{e}}(\nu)\bigr)
                \]
                \State \Return $\hat{\mathbf{e}}^\star=\hat{\mathbf{e}}(\nu^\star)$, $\nu^\star$
            \EndIf
        
            \State $\mathcal{B}^{(t+1)} \gets \operatorname{Top}_W\bigl(\operatorname{pc}^{(t+1)},\mathcal{P}\bigr)$
        \EndFor

        \If{$\mathcal{B}^{(T_{\max})}\neq \emptyset$}
            \State $v^\star \gets \arg\max_{v \in \mathcal{B}^{(T_{\max})}} P(v)$
            \State \Return $\hat{\mathbf{e}}^\star=\hat{\mathbf{e}}(v^\star)$, $v^\star$
        \Else
            \State \Return failure
        \EndIf

    \end{algorithmic}
\end{algorithm}

\section{Simulation Results}
\label{sec:performance}
We simulate the performance of the multistage decoder for the family of BB codes and the LP Tanner code. Unless noted otherwise, the maximum number of iterations for nMS is set to $100$ and every decoder shown in the plots follows parallel scheduling. The normalization parameter has been set to $0.875$. The nMS-OSD-$10$ decoder is considered the benchmark for evaluating the performance of multistage decoder and the nMS-OSD-$10$ decoder were generated based on~\cite{window}. A BSC with crossover probability $\alpha$ resulting in the bit-flip ($X$) errors is assumed for the noise model.

\begin{figure}[htbp]
    \centering
\includegraphics[width=0.5\textwidth]{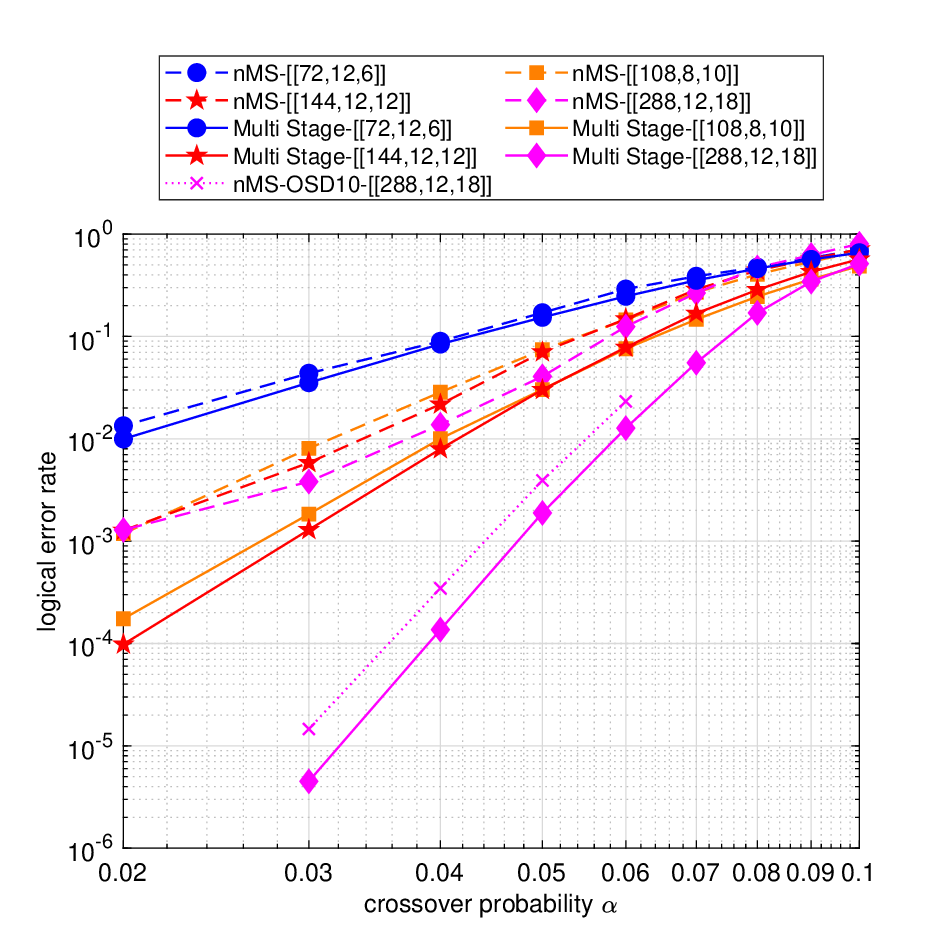}
    \caption{Multistage decoder performance for BB codes of various distances. The maximum number of stages is $11$, beam width is $64$, and top-$K$ selection parameter is $1$. The maximum iteration number is set to $L=100$.}
    \label{fig:BBVariousDistances}
\end{figure}

\begin{figure}[htbp]
    \centering
\includegraphics[width=0.5\textwidth]{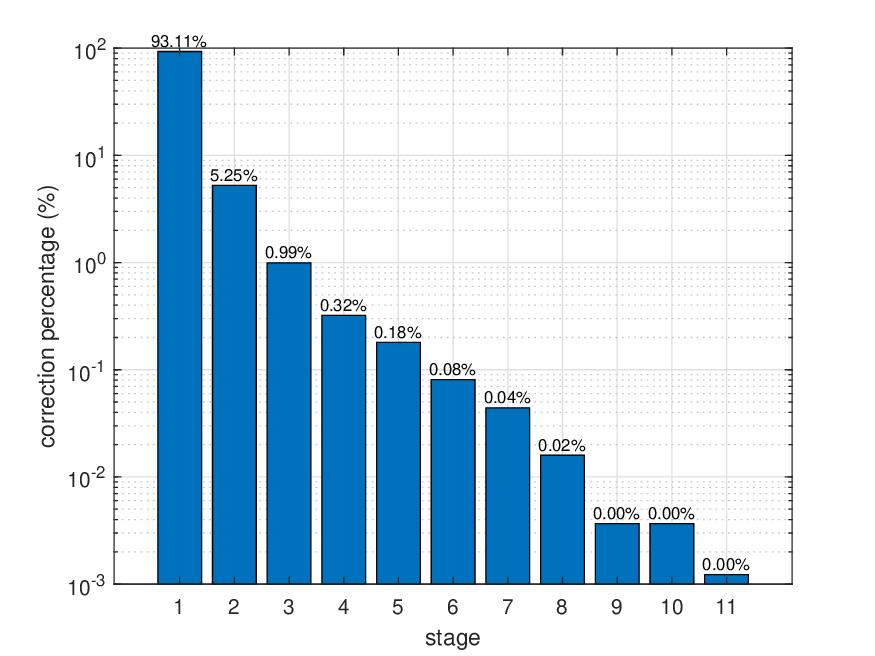}
    \caption{Stage correction of the multistage decoding for BB-$[[288, 12, 18]]$ at $\alpha = 0.03$. The maximum number of stages is $11$, beam width is $64$, and top-$K$ selection parameter is $1$.}
    \label{fig:BB288CorrectionStage}
\end{figure}

\begin{figure}[htbp]
    \centering
\includegraphics[width=0.5\textwidth]{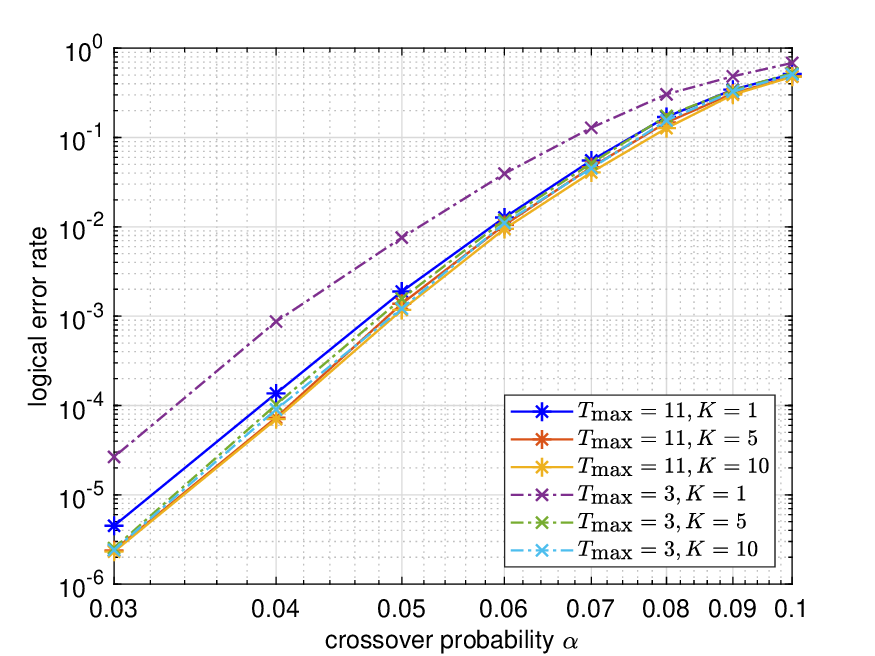}
    \caption{Multistage decoder performance for BB-$[[288, 12, 18]]$ codes for different maximum number of stages and top-$K$ selection parameter. The maximum iteration number is set to $L=100$.}
    \label{fig:DifferingBeamsearchBB}
\end{figure}

\begin{figure}[htbp]
    \centering
\includegraphics[width=0.5\textwidth]{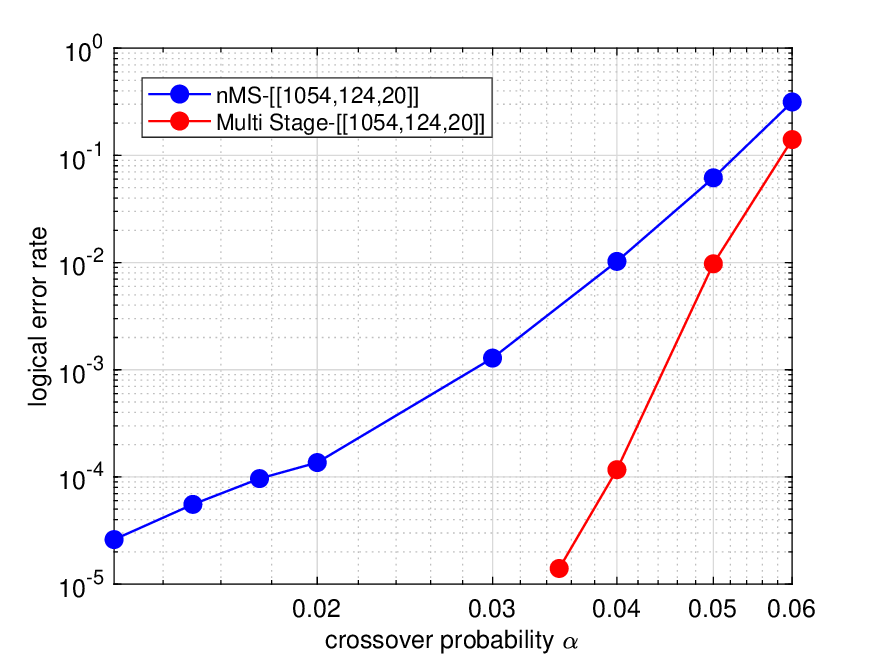}
    \caption{Multistage decoder performance for the $[[1054, 124, 20]]$ LP Tanner code. The maximum number of stages is $11$, beam width is $64$, and top-$K$ selection parameter is $1$. The maximum iteration number is set to $L=100$.}
    \label{fig:LPTanner}
\end{figure}

\begin{figure}[htbp]
    \centering
\includegraphics[width=0.5\textwidth]{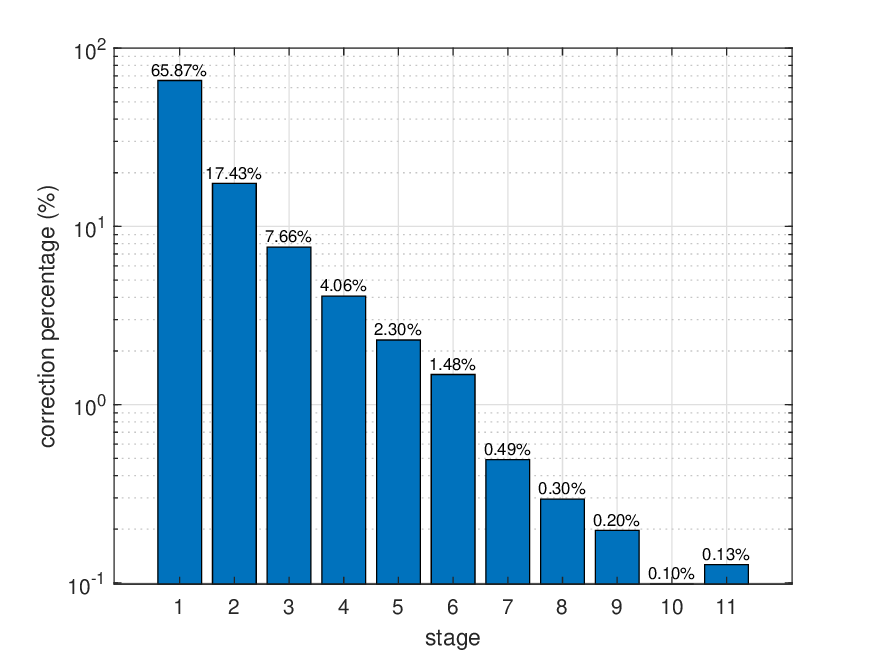}
    \caption{Stage correction of the multistage decoding for LP Tanner code at $\alpha = 0.04$. The maximum number of stages is $11$, beam width is $64$, and top-$K$ selection parameter is $1$.}
    \label{fig:LPTannerStageCorrection}
\end{figure}

\begin{figure}[htbp]
    \centering
\includegraphics[width=0.5\textwidth]{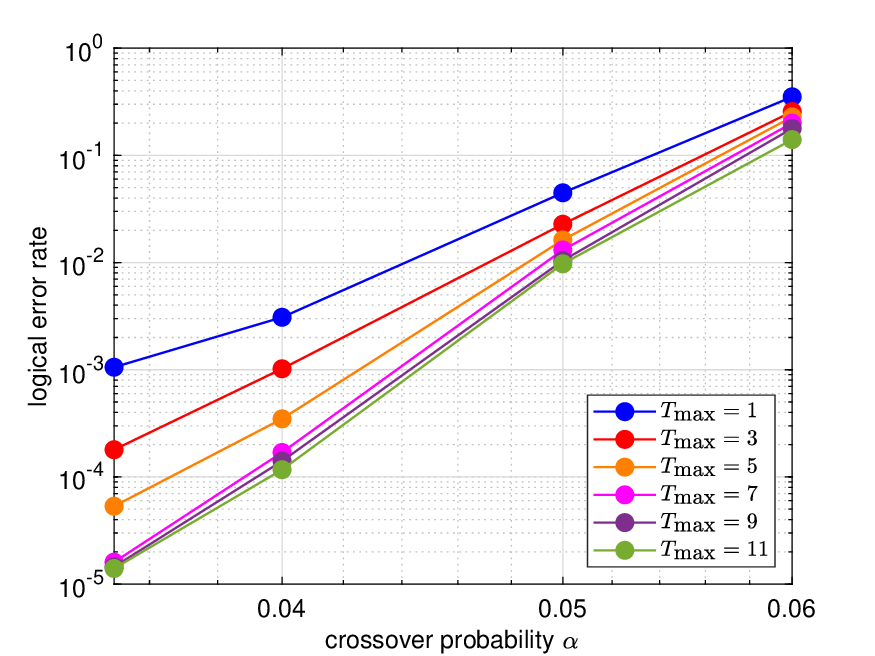}
    \caption{Multistage decoder performance for LP Tanner code for varying maximum number of stages. The maximum number of stages is set to $1, 3, 5, 7, 9, 11$. The maximum iteration number is set to $L=100$.}
    \label{fig:DifferentStagesBeamsearchLPTanner}
\end{figure}

Fig.~\ref{fig:BBVariousDistances} illustrates the logical error rate (LER) performance of nMS, nMS-OSD-$10$, and the proposed multistage decoder for several BB code instances. For BB codes, the dominant uncorrectable errors by nMS are degenerate errors supported on symmetric stabilizers, as provided in~\cite{DimitrisEnhanced}. The proposed multistage decoder is able to correct not only these dominant nMS failure events, but also classical trapping set configurations inherent to the code structure. For benchmarking purposes, we also compare the proposed multistage decoder with nMS-OSD-$10$ for BB code of length $288$. The proposed method achieves notable gains. In particular, for BB-$288$ at $\alpha = 0.03$, it provides a $286 \times$ improvement over nMS and $3.2\times$ over nMS-OSD-$10$. In the simulations, the selected values of $c_U$, $c_E$, and $c_O$ are $0.5$, $0.3$, and $0.2$, respectively.

Fig.~\ref{fig:BB288CorrectionStage} shows the percentage of successful corrections occurring at each stage for BB-$288$ code under the proposed multistage decoder at crossover probability $\alpha = 0.03$. The decoder is configured with a maximum search depth of $11$, beam width $64$, and top-$K$ selection parameter $1$. As illustrated in the Fig.~\ref{fig:BB288CorrectionStage}, most decoding failures are corrected in the first few stages, whereas only a small fraction of error events require correction at deeper stages. This stage correction behavior demonstrates that the proposed search strategy is effective in identifying correction early in the decoding process, which in turn supports the practical feasibility of the multistage framework.

In Fig.~\ref{fig:DifferingBeamsearchBB}, we describe the effect of different parameters of the proposed multistage decoder, with the beam width fixed at 64. In particular, we study the impact of the maximum number of stages $T_{\textrm{max}}$ and the top-$K$ selection parameters. First, with $T_{\textrm{max}}=11$ and $K$ varying from $1$ to $5$, we observe $1.9\times$ improvement in LER at crossover probability $\alpha=0.03$. Increasing $K$ further from $5$ to $10$ yields only negligible additional improvement, indicating that the performance has essentially saturated. As shown in Fig.~\ref{fig:BB288CorrectionStage}, approximately $99.35\%$ of all successful decodings occur within the first three stages. Motivated by this observation, rather than allowing the maximum number of stages to remain as large as $11$, we reduce it to $T_{\textrm{max}} = 3$ and increase the top-$K$ selection parameter in an attempt to compensate for the reduced search depth. Under this setting, reducing the maximum number of stages to $3$ while keeping $K=1$ results in a significant performance degradation. However, increasing the top-$K$ parameter to $K=5$ substantially mitigates this loss, yielding a $10.5\times$ improvement at crossover probability $\alpha = 0.03$. Increasing $K$ further to $10$ again provides only marginal additional gain compared to the case $K=5$. We compare the best performing parameter setting, $T_{\textrm{max}} = 11$ and $K=10$, with the reduced depth of search $T_{\textrm{max}} = 3$ and $K=10$. The comparison shows that the former achieves only a negligible additional gain, a $1.04\times$ improvement at crossover probability $\alpha = 0.03$ and $1.29\times$ improvement at crossover probability $\alpha = 0.04$. The result suggest that most of the performance benefit can be obtained with a smaller search depth, provided that a sufficiently large top-$K$ parameter is used.

In Fig.~\ref{fig:LPTanner}, we compare the decoding performance of the $[[1054, 124, 20]]$ LP Tanner code  under nMS and the proposed multistage decoding scheme. For the multistage decoder, the maximum number of stages is set to $11$, the beam width is chosen as $64$, and the top-$K$ selection parameter is fixed to $1$. The LP Tanner code contains both classical and quantum trapping set configurations, and the multistage decoder is capable of mitigating failure events arising from both. As can be seen, at crossover probability $\alpha = 0.4$, the multistage decoder provides a two order of magnitude improvement over conventional nMS decoding.

Fig.~\ref{fig:LPTannerStageCorrection} illustrates the stage wise distribution of successful corrections for the LP Tanner code under the proposed multistage decoder at crossover probability $\alpha = 0.04$. The decoder is configured with a maximum search depth of $11$, beam width $64$, and top-$K$ selection parameter $1$. As in the BB-$288$, Fig.~\ref{fig:BB288CorrectionStage}, most of the error patterns that are uncorrectable by conventional nMS are resolved in the early stages of the search. However, a comparison between the stage wise correction statistics of BB-$288$ and LP Tanner-$1054$ reveals a higher correction percentage in the first stage. This observation suggests that, for the LP Tanner code, a greater number of stages are required to resolve a larger portion of the failure error patterns. Overall, these results indicate that the stage-wise correction statistics depend strongly on the underlying code structure and, in particular, on the nature of its trapping set configurations. In Fig.~\ref{fig:DifferentStagesBeamsearchLPTanner}, we show the effect of the maximum number of stages, with $T_{\textrm{max}}=\{1,3,5,7,9,11\}$, while fixing the beamwidth to $64$, and the Top-$K$ selection parameter to $1$. As observed in the Fig.~\ref{fig:DifferentStagesBeamsearchLPTanner}, increasing the number of stages improves the successful decoding. However, beyond a certain search depth, the improvement becomes smaller, indicating a saturation effect in performance.

The results obtained for the BB codes achieve performance comparable to that of BP-OSD-$10$, and in some cases outperform it. We also investigate the effect of different parameter settings in the proposed multistage decoder, with particular emphasis on the maximum number of stages and the top-$K$ selection parameter, in order to obtain improved trade offs between decoding latency and performance.

\section{Conclusions and Future Work}
We have proposed multistage decoding scheme that iteratively applies the BP algorithm. Based on the proposed heuristic metric, the decoder identifies unreliable bits and forces them to specific values, thereby enabling the decoding process to resolve trapping sets and reach to the attractor. More precisely, the method constructs a search tree and, at each stage, forces the values of the variable nodes with the highest unreliability scores. In addition, we introduce a pruning metric to concentrate the search on the most promising nodes. For the BB codes, the proposed decoder achieves performance comparable to that of BP-OSD-$10$, and in some cases outperforms it. In addition, for the LP Tanner code, the proposed method improves the decoding performance by two orders of magnitude at crossover probability $\alpha=0.04$ compared to the conventional nMs.

\IEEEtriggeratref{25}
\bibliographystyle{IEEEtran}
\bibliography{bib/totalRefs.bib}
\end{document}